\documentclass[12pt,a4paper]{article}
\usepackage{amsmath}
\usepackage{amssymb}
\usepackage{amsxtra}
\begin{document}
\newtheorem{theorem}{Theorem}
\newtheorem{axiom}{Axiom}
\newtheorem{corollary}{Corollary}
\author{Ilja Schmelzer\thanks
       {WIAS Berlin}}

\title{A Lattice Fermion Doublet With A Generalization Of The
Ginsparg-Wilson Relation} \sloppypar\sloppy

\maketitle
\begin{abstract}

We present a new staggered discretization of the Dirac operator.  In
comparison with standard staggered fermions, real and imaginary parts
are located in different nodes.  Doubling gives only a doublet of
Dirac fermions which we propose to interpret as a physical (lepton or
quark) doublet.  Contrary to usual staggered fermions, we have no
exact chiral symmetry but obtain a generalization of the
Ginsparg-Wilson relation.

\end{abstract}

\section{Introduction}

In this paper we present a new discretization of the Dirac equation.
In comparison with staggered fermions \cite{Kogut} it creates not four
but only two flavours of Dirac fermions.  This has been reached by
placing not only different spin components, but also their real and
imaginary parts into different nodes.  These sixteen steps of freedom
(two fermions) can be understood, in some sense, as the result of
``doubling'' of a real scalar step of freedom $\varphi(n)$ on the
lattice.

Moreover, these two fermions live on different sub-lattices ($\psi_o$
on ``odd'' nodes, $\psi_e$ on ``even'' nodes), thus, we obtain also a
single fermion (eight steps of freedom) on the lattice by omitting
half of the lattice nodes.  But, if the neutrino is a Dirac particle,
all fermions appear in doublets of Dirac particles.  In this context,
the appearance of a fermion doublet in this discretization may be not
a bug but a feature which allows to explain why Dirac particles appear
in such doublets.

In our approach the complex structure is an operator among others.
Moreover, there is no natural complex structure, but, instead, a
quaternionic structure.  The nature of this structure is a topic of
future research.

Once a complex structure is defined, we have to define the operator
$\gamma^5$ on the grid.  Now, there is no exact chiral symmetry.
Instead, we obtain formulas for some operators $\gamma^5$ on the grid
which define a generalization of the Ginsparg-Wilson relation.

\section{A Real Representation Of The Dirac Algebra}

We forget -- for some time -- about the complex structure.  Instead of
the usual representations with four complex fields, we use an
eight-dimensional real representation of the operators $\gamma^\mu$
defined here by their linear combination with $\partial_\mu$:

\begin{equation}
\label{spinor}
\gamma^0\partial_0-\gamma^i\partial_i =_{def}
\left(
\begin{array}{rrrrrrrr}
 \partial_0& \partial_1& \partial_2&           & \partial_3&&&\\
-\partial_1&-\partial_0&           & \partial_2&& \partial_3&&\\
-\partial_2&           &-\partial_0&-\partial_1&&& \partial_3&\\
           &-\partial_2& \partial_1& \partial_0&&&& \partial_3\\           
-\partial_3&&&&-\partial_0&-\partial_1&-\partial_2&           \\
&-\partial_3&&& \partial_1& \partial_0&           &-\partial_2\\
&&-\partial_3&& \partial_2&           & \partial_0& \partial_1\\
&&&-\partial_3&           & \partial_2&-\partial_1&-\partial_0          
\end{array}
\right)
\end{equation}

In the context of this representation, it seems also natural to define
(by their linear combination with scalar parameters $m_i$) the
following operators $\beta^i$:

\begin{equation}
\label{mass_matrix}
m_i\beta^i =_{def}
\left(
\begin{array}{rrrrrrrr}
    & m_1& m_2&           & m_3&&&\\
 m_1&    &           & m_2&& m_3&&\\
 m_2&           &    &-m_1&&& m_3&\\
           & m_2&-m_1&    &&&& m_3\\           
 m_3&&&&    &-m_1&-m_2&           \\
& m_3&&&-m_1&    &           &-m_2\\
&& m_3&&-m_2&           &    & m_1\\
&&& m_3&           &-m_2& m_1&              
\end{array}
\right)
\end{equation}

The following operator equation holds:

\begin{equation}
(\gamma^0\partial_0-\gamma^i\partial_i + m_i\beta^i)^2 
 = -\square + \delta^{ij} m_i m_j
\end{equation}

This can be easily seen -- this operator iterates three times, in each
coordinate direction, the same trick:\footnote{This observation also
suggests how to iterate this construction to arbitrary dimension.}

\begin{equation}
\left(
\begin{array}{cc}
A&(m_i+\partial_i)I\\(m_i-\partial_i)I&-A
\end{array}
\right)^2
=
(A^2+(m_i+\partial_i)(m_i-\partial_i)I)\left(
\begin{array}{cc}
I&0\\0&I
\end{array}
\right)
\end{equation}

As follows immediately, the $\gamma^\mu$ define a representation of
the Dirac matrices, and the matrices $\beta^i$ fulfil the following
anticommutation relations:

\begin{equation}
\beta^i\beta^j+\beta^j\beta^i=\delta^{ij}
\end{equation}

and anticommute with all $\gamma^\mu$:

\begin{equation}
\beta^i\gamma^\mu+\gamma^\mu\beta^i=0
\end{equation}

It is also easy to see (and to generalize to arbitrary dimension) that

\begin{equation}
\gamma^0(\gamma^1\beta^1)(\gamma^2\beta^2)(\gamma^3\beta^3)=1.
\end{equation}

\section{Complex Structures}

In this representation we have not yet defined a complex structure.
There are several things in the standard approach to Dirac fermions
which depend on them.

First, in the standard approach a Hermitean scalar product
$\langle.,.\rangle$ is widely used.  We have only a standard Euclidean
scalar product $(.,.)$ yet.  Now, for a complex structure $i$ these
notions are closely related in a simple way:

The Hermitean scalar product defines an Euclidean scalar product by

\begin{equation}
(\psi,\phi)=\frac{1}{2}(\langle\psi,\phi\rangle+\langle\phi,\psi\rangle
\end{equation}

so that $(i\psi,i\phi)=(\psi,\phi)$.  For a complex structure $i$,
$i^2=-1$, with this property this Hermitean scalar product is defined
by the Euclidean scalar product by defines the Hermitean scalar
product

\begin{equation}
\langle\psi,\phi\rangle=(\psi,\phi)-i(\psi,i\phi).
\end{equation}

Thus, we should not care about the Hermitean scalar product, the
Euclidean scalar product is all we need.  Note also that for a complex
linear operator $A$, which is a real linear operator with $[A,i]=0$,
the Hermitean adjoint operator $A^+$ and the Euclidean adjoint
operator $A^*$ cooinside: 
$\langle A^*\psi,\phi\rangle=\langle\psi,A\phi\rangle$.
As a consequence, the classical properties of the
$\gamma$-matrices 

\begin{equation}
(\gamma^\mu)^+ = \gamma^0\gamma^\mu\gamma^0
\end{equation}

are equivalent to

\begin{equation}
(\gamma^\mu)^* = \gamma^0\gamma^\mu\gamma^0.
\end{equation}

These properties are fulfilled in our representation for the standard
Euclidean scalar product $(.,.)$ in $\mathbb{R}^8$.

Next, the ``classical'' operator $\gamma^5=
-i\gamma^0\gamma^1\gamma^2\gamma^3$ also depends on the complex
structure.  A natural replacement which does not depend on it -- the
expression $\gamma^0\gamma^1\gamma^2\gamma^3$ -- we denote with
$\iota$:

\begin{equation}
\iota=_{def}\gamma^0\gamma^1\gamma^2\gamma^3=\beta^1\beta^2\beta^3 \hspace{1cm}
\iota\gamma^\mu+\gamma^\mu\iota=0\hspace{1cm}
(\iota)^2 = -1
\end{equation}

But, of course, we have to introduce a complex structure if we want to
connect the Dirac fermion in the usual way with gauge fields.  The
properties we need for a complex structure $i$ are $i^{-1}=i^*=-i$ and
$[\gamma^\mu,i]=0$.  Now, an interesting point is that there are
several candidates for such a structure:

\begin{eqnarray}
i &= \beta^1\beta^2 &= \iota\beta^3\\
j &= \beta^2\beta^3 &= \iota\beta^1\\
k &= \beta^3\beta^1 &= \iota\beta^2
\end{eqnarray}

which together define a quaternionic structure:\footnote{The classical
representation $ij=k$ can be obtained using reverse signs for $i,j,k$,
but we prefer this sign convention because it gives
$\gamma^5=\beta^3$.}

\begin{equation}
i j = - j i = -k;\; 
j k = - k j = -i;\; 
k i = - i k = -j;\; 
i^2 = j^2 = k^2 = -1
\end{equation}

For each candidate $i$ for a complex structure, we obtain an own
operator $\gamma^5=_{def}-i\iota$.  Especially for $i=\iota\beta^3$ we
obtain $\gamma^5=-\iota\beta^3\iota=\beta^3$.  Thus, it seems that to
fix the complex structure we somehow have to break spatial symmetry,
prefer one spatial direction.

\section{Discretization Of The Dirac Equation}

Our representation is appropriate for a discretization of the Dirac
equation on a regular hyper-cubic lattice.  It can be obtained in a
quite simple way: We start with a naive central difference
approximation

\begin{equation}
\partial_i\psi(n)\to\frac{1}{2a_i}(\psi(n+a_i)-\psi(n-a_i)).
\end{equation}

This naive discretization leads to the problem of ``fermion
doubling''.  The continuous limit of this set of discrete equations
gives not only the original Dirac equation, but also additional,
highly oscillating components, the ``doublers''.  In classical
computations such doublers may be often ignored, but in quantum
computations, where the number of steps of freedom is important (Pauli
principle) this is no longer possible.  We obtain in each direction a
factor to, thus, $2^4=16$ doublers.  Fortunately, eight pairs of
doublers decouple in a really simple way: It is sufficient to hold
only one of the eight real components $\psi^a$ per node.  On the
three-dimensional reference cube
$(\varepsilon_1,\varepsilon_2,\varepsilon_3)$, $\varepsilon_i\in\{0,1\}$
we obtain the following locations for the eight components:

\begin{equation}
\label{location3D}
\begin{array}{@{\psi}l@{\mbox{ located at }(}c@{),\hspace{1cm} \psi}l@{\mbox{ located at }(}c@{);}}
^0&0,0,0&^4&0,0,1\\
^1&1,0,0&^5&1,0,1\\
^2&0,1,0&^6&0,1,1\\
^3&1,1,0&^7&1,1,1
\end{array}
\end{equation}

What remains are sixteen steps of freedom (eight steps of freedom on
two time steps which we need because of our use of central differences
in time) which corresponds to a doublet of Dirac fermions\footnote{
This is a variant of a well-known approach to solve the ``fermion
doubling'' problem -- so-called staggered fermions \cite{Kogut}.  The
standard staggered grid approach reduces the doublers only by a factor
four.  Because we ignore the complex structure of the standard
representation, we are free to place ``real'' and ``imaginary'' part
of the complex fields into different nodes. This gives the additional
reduction by factor two.}.  Note that our discretization may be
interpreted as a way to discretize the d'Alembert equation for a
single scalar step of freedom $\varphi(n)$ with central differences,
which gives $2^4 = 16$ doublers.

Now, the last doublet decouples too, but in a slightly less trivial
way: We can distinguish ``even'' and ``odd'' nodes on the full
space-time lattice.  The central difference equations on even (odd)
nodes connects only values on odd (even) nodes.  Thus, we obtain two
fermions $\psi_e$ and $\psi_o$ on even resp. odd nodes.  On the
four-dimensional reference cube
$(\varepsilon_0,\varepsilon_1,\varepsilon_2\varepsilon_3),
\varepsilon_i\in\{0,1\}$ we have

\begin{equation}
\begin{array}{@{\psi_e}l@{\mbox{ located at }(}c@{),\hspace{1cm} \psi_e}l@{\mbox{ located at }(}c@{);}}
^0&0,0,0,0&^4&1,0,0,1\\
^1&1,1,0,0&^5&0,1,0,1\\
^2&1,0,1,0&^6&0,0,1,1\\
^3&0,1,1,0&^7&1,1,1,1
\end{array}
\end{equation}

\begin{equation}
\begin{array}{@{\psi_o}l@{\mbox{ located at }(}c@{),\hspace{1cm} \psi_o}l@{\mbox{ located at }(}c@{);}}
^0&1,0,0,0&^4&0,0,0,1\\
^1&0,1,0,0&^5&1,1,0,1\\
^2&0,0,1,0&^6&1,0,1,1\\
^3&1,1,1,0&^7&0,1,1,1
\end{array}
\end{equation}

But, instead of removing one sub-mesh to describe a single Dirac
fermion, we propose to accept above doublers as a way to describe a
physically meaningful flavour doublet.  Remarkably, if the neutrino is
a standard Dirac particle, then all fermions of the standard model
appear in doublets.  The appearance of a fermion doublet in our
approach may be, therefore, not a bug but a feature which allows to
explain the existence of these doublets.

\subsection{Lagrange Formalism}

To define the Lagrange formalism on the lattice note that the
equations for the $\psi_o$ are located at the nodes of $\psi_e$ and
reverse.  Therefore, we can use the Lagrangian

\begin{equation}
L = \sum \psi_e (D \psi_o) = - \sum \psi_o (D \psi_e)
\end{equation}

and obtain the equation for $D \psi_{o/e} = 0$ as the Euler-Lagrange
equation for $\psi_{e/o}$.  We can also rewrite this Lagrangian as

\begin{equation}
L = \frac{1}{2} \sum \varphi (-1)^{\epsilon_0}\gamma^0 (D \varphi)
\end{equation}

\subsection{Fermion Families and Lattice Distortions}

We have observed that to fix the complex structure we somehow have to
break spatial symmetry.  But there is a simple way out of this.
Instead of one scalar step of freedom $\varphi(n)$ on the lattice, we
can consider a vector field -- thus, three components $\varphi^i(n)$.
Now, each component has a natural ``preferred direction'' and,
therefore, a natural complex structure.

Moreover, a vector on a lattice is a quite natural step of freedom.
It is, for example, the natural way to describe lattice distortions
with a shift vector field $u^i(n)$.

On the other hand, in the standard model we have three fermion
families -- three copies of each fermion.  This suggests to explain on
the kinematic level the three fermion families using the hypothesis
that the fundamental steps of freedom of the ``theory of everything''
are three-dimensional vector fields $u^i(n)$ of spatial distortions.

\section{Chiral Symmetry On The Lattice}

One problem with standard staggered fermions \cite{Kogut} is that they
have the wrong number of doublers (four) to allow a natural physical
interpretation in the standard model.  In our approach, we have only a
pair of Dirac fermions, and pairs of fermions appear in the standard
model as quark pairs as well as lepton pairs (if the neutrino is a
Dirac particle).

The other problem is that there is exact chiral $\gamma^5$ symmetry on
the lattice.  As a consequence, the doublers appear in pairs with
reverse chiral charge.  This does not fit the situation in the
standard model (cf. \cite{Gupta}).  Now, in our approach we do not
have exact chiral $\gamma^5$ symmetry.  Instead, we have a replacement
for this symmetry.  This replacement fulfills properties which define
a generalization of the famous Ginsparg-Wilson (GW) relation
\cite{Ginsparg}.

Let's assume now that one of the complex structures, namely
$i=\iota\beta^3$, has been chosen.  To understand chiral symmetry we
have to define the operator $\gamma^5=\beta^3$ on the lattice.  Note
that it cannot be anymore a pointwise operator as for Wilson fermions
and staggered fermions -- it connects components which are located in
different points.  Now, we propose to consider the following operator
as a candidate for $\gamma^5$ on the lattice:

\begin{equation}
(\gamma^5 \phi)(n_{even}) = \phi(n_{even}-h_z),\hspace{1cm}
(\gamma^5 \phi)(n_{odd }) = \phi(n_{odd }+h_z)
\end{equation}

It is easy to see that it approximates the continuous $\gamma^5$.
More interesting is that some exact properties remain valid:

\begin{equation}
(\gamma^5)^*=\gamma^5,\hspace{1cm}  (\gamma^5)^2=1
\end{equation}

We can also define, as an alternative, the operator $\tilde{\gamma}^5$
by

\begin{equation}
(\tilde{\gamma}^5 \phi)(n_{even}) = \phi(n_{even}+h_z),\hspace{1cm}
(\tilde{\gamma}^5 \phi)(n_{odd }) = \phi(n_{odd }-h_z)
\end{equation}

Similarly, we obtain

\begin{equation}
(\tilde{\gamma}^5)^*=\tilde{\gamma}^5,\hspace{1cm}  (\tilde{\gamma}^5)^2=1
\end{equation}

If we define the operators $V,O$ by
$ \tilde{\gamma}^5 =  \gamma^5 V = \gamma^5(1-aO) $

we obtain the Ginsparg-Wilson (GW) relation for $O$:

\begin{equation}
O \gamma^5 + \gamma^5 O = a O \gamma^5 O
\end{equation}

Moreover, we have also the following important commutation properties
with $D$

\begin{equation}
\tilde{\gamma}^5 D + D \gamma^5 = 0,\hspace{1cm}
\gamma^5 D + D \tilde{\gamma}^5 = 0,
\end{equation}

\begin{equation}
V D - D V = 0,\hspace{1cm}
O D - D O = 0.
\end{equation}

This allows to define two sets of chiral projector operators

\begin{equation}
\tilde{P}_\pm = \frac{1}{2} (1\pm\tilde{\gamma}^5),\hspace{1cm}
      {P}_\pm = \frac{1}{2} (1\pm      {\gamma}^5).
\end{equation}

Similar pairs of projectors play a central role in approaches to
chiral gauge theory based on the GW relation (\cite{Golterman},
\cite{Luescher}) and it's generalizations (\cite{Kerler}) as domain
wall fermions \cite{Shamir}, Neuberger's overlap operator
\cite{Neuberger}, proposals by Fujikawa \cite{Fujikawa} and Chiu
\cite{Chiu}.

On the other hand, it should be noted that our approach does not fit
exactly into these schemes.  The operators $V, O$ do not have the
spectral properties of the similar operators considered, for example,
by \cite{Golterman},\cite{Kerler}.  It seems that these differences
may be understood as caused by different aims.  The aim of the
standard GW approach is to obtain a single Weyl fermion on the
lattice, without any doublers.  In our approach we allow some doubling
which gives a pair of Dirac fermions on the lattice, but we want
nontrivial chiral symmetry to allow nontrivial chiral interactions
between the doublers.

If the properties of $\gamma^5, \tilde{\gamma}^5$ we have found are
sufficient to develop a consistent chiral gauge theory on this lattice
remains to be shown.  Note that the situation is further complicated
by the non-pointwise character of the complex structure which remains
to be understood.

\end{document}